\documentclass[aps,prl,
preprint,
showpacs,superscriptaddress,groupedaddress,sort&compress,numbers]{revtex4-1} 

\usepackage{graphicx}  
\usepackage{dcolumn} 
\usepackage{bm}        
\usepackage{amssymb}  
\usepackage{siunitx}
\usepackage{placeins}
\usepackage{natbib}

\usepackage{amsmath}

\usepackage[utf8]{inputenc}

\begin{document}

\widetext

\title{Skyrmion meets magnetic tunnel junction:\\ an efficient way for electrical skyrmion
detection investigated by ab initio theory}

\author{Jonas Friedrich Sch{\"a}fer-Richarz\textsuperscript{*}}
\author{Philipp Risius} 
\author{Michael Czerner} 
\author{Christian Heiliger}

\vskip 0.25cm
\affiliation{Institute for Theoretical Physics, Justus Liebig University Giessen, 35392 Giessen, Germany}           
\affiliation{Center for Materials Research (LaMa), Justus Liebig University Giessen, 35392 Giessen, Germany}                  
                             
\date{\today}

\begin{abstract}

In our proof-of-principle study we examine the influence of skyrmions on magnetoresistive transport. In particular, we show that magnetic tunnel junctions are a technologically appealing and promising way for electrical detection of non-collinear magnetic structures. The calculated effect is shown to originate from scattering between different k-states and cannot be identified through densities of states alone. Our results suggest that the detection efficiency strongly depends on the utilized materials. 
\end{abstract}

\maketitle

Magnetic skyrmions are swirl-like magnetic structures which are stabilized by a complex interplay of various competing magnetic interactions, such as the Dzyaloshinskii-Moriya interaction (DMI) \cite{roess1,SkyrLattice,nagaosaOrig,dmi,moriya,skyrme}. While they can arise in noncentrosymmetric bulk materials such as MnSi \cite{SkyrLattice,tokunaga}, thin magnetic films were found to be practical hosts for skyrmions, which allows more flexibility in their design \cite{kis, dupe, symmbreaksurf, woo, Soumy, Jaiswal, filmHeinze,boulle,hoffmann,dovzh,zhang}. Since skyrmions have been discovered, concepts for novel electrical devices were discussed.  The possibility to move, create, and delete skyrmions combined with their topological stability lead to concepts such as skyrmion based racetrack memory and even logical devices \cite{buettner,iwasaki,yu,writedelete,fert,sampaio,parkin,ZhangLogic,fert17,sitte,hsu,tomasello,magnondrive,jiang,loreto2018creation}. One essential ingredient for the realization of these approaches is the ability to detect skyrmions with minimal effort and minimal disturbance of the skyrmion. 

For applications an effective and preferably electrical detection of skyrmions is required. In experiments scanning tunneling microscopy (STM) may be used to detect non-collinear structures such as skyrmions. With this technique, non-collinear transport effects have been demonstrated and were called either non-collinear magnetoresistance (NCMR) or tunnelling spin-mixing magnetoresistance (TXMR) \cite{lounis, ncmr, hann}. A major drawback of this method is that STM based devices are impractical for large-scale production and application. Similar to references \cite{lounis, bode, ncmr, hann}, it is our aim to detect the non-collinear magnetic texture of a skyrmion via tunneling currents. However, we employ magnetic tunnel junctions (MTJ) instead of STM for detection. MTJs can be fabricated by lithography techniques and are thus much better suited for industrial applications.

An important effect to measure relative magnetization directions in MTJs is tunnel magnetoresistance (TMR) \cite{moodera}. In typical MTJs, two ferromagnetic layers are separated by a thin insulating layer. The tunneling probability through the junction then depends on the relative angle between the magnetization directions. In case of amorphous barriers, the Jullière model \cite{julliere1975tunneling} explains TMR in terms of spin-polarization of the density of states, with limited effect sizes of typically less than 100$\%$ \cite{heiliger2006microscopic}. With crystalline MgO barriers, however, transport is coherent and $k_\parallel$-conserving, leading to experimentally achieved effect sizes of several hundred percent  \cite{lee2007effect}. This increase relies on a high spin-polarization per $k$-point, which is the relevant quantity in case of coherent tunneling \cite{Gradhand}, and symmetry filtering in MgO \cite{butler2008tunneling}. Symmetry filtering in MgO means that electronic states with $\Delta_1$-symmetry can tunnel much better through MgO than other states. In iron $\Delta_1$-states only occur in the majority channel. This together with similar filtering effects at other $k$-points is responsible for the large TMR ratios.

While detecting skyrmions using the TMR effect might seem straightforward, the necessary magnetic reference layer must be expected to disturb the skyrmion \cite{zhang2018skyrmions}. The related tunneling anisotropic magnetoresistance (TAMR) effect \cite{gouldtamr,matostamr,khantamr}, however, does not require a magnetic reference layer. This effect only depends on the magnetization direction of one ferromagnetic layer, is caused by spin-orbit coupling (SOC), and is typically rather small compared to TMR \cite{matostamr}. 

We predict much larger effects, compared to STM studies, for an MgO based MTJ using vanadium and copper as non-magnetic leads. A deeper analysis reveals that the nature of the underlying effect is based on $k$-scattering together with the symmetry filtering property of MgO. Since the presence of the skyrmion opens new, dominant transport channels, the structure is essentially acting like a spin valve with a single magnetic layer. Similar to TMR \cite{butler2008tunneling,heiliger2006microscopic}, the mechanism goes beyond the analysis of spin-polarized densities of states as used in references \cite{lounis, bode, ncmr, hann}. While phenomenologically the effect is the same as TXMR or NCMR, the mechanism is different and leads to a much higher magnetoresistance ratio. 

\begin{figure}
\includegraphics[scale=0.3]{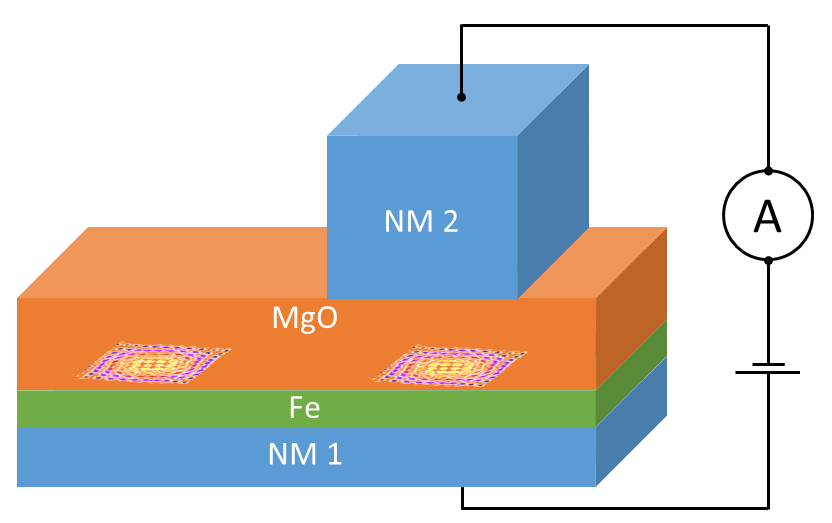}
\caption{\label{fig:geofig} The geometrical structure of the system. The skyrmions are located in the xy-plane in the thin iron monolayer. The presence of a skyrmion in the MTJ alters the conductance and can thus be detected.} 
\end{figure}

Our work is an ab initio based proof-of-principle study. To describe the system, we are using an atomic monolayer of iron in which the skyrmion is located. This iron layer is the sole magnetic layer and is sandwiched between a non-magnetic lead (NM1) and the insulating MgO barrier (see FIG. \ref{fig:geofig}). The latter consists of 3 monolayers and is covered by the second non-magnetic lead (NM2). To examine the influence of the lead material, we compare three systems, which differ purely in the materials choice for the non-magnetic leads, using either copper or vanadium. 
As in previous works \cite{artCu}, 
the non-magnetic leads are modeled by a bcc structure with a lattice constant of 2.866 $ \si{\angstrom}$ while out-of-plane lattice relaxation is taken into account for the layers directly adjacent to the MgO.

\begin{figure}
\includegraphics[scale=0.22]{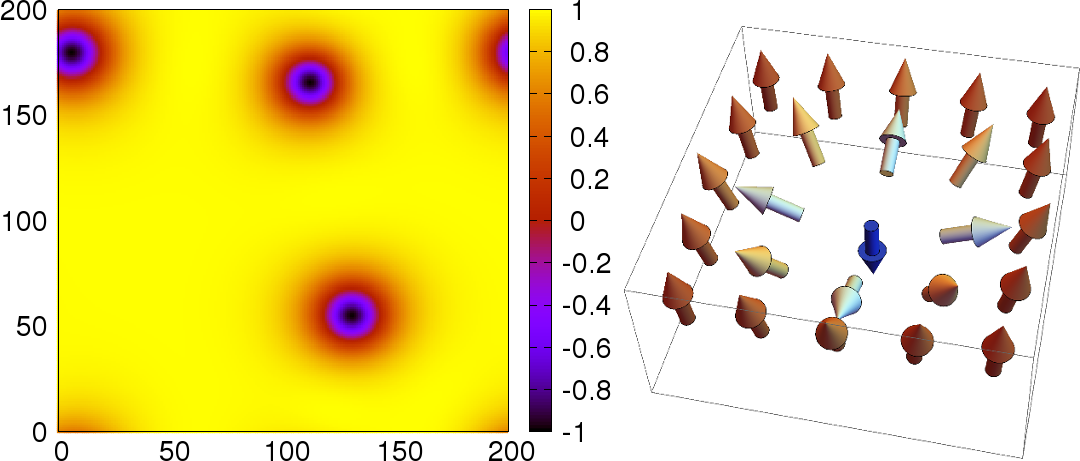}
\caption{\label{fig:nskyr} Left: Skyrmions in our copper based MTJ from atomistic spin dynamics simulations. The entire frame has a size of 200$\times$200 atoms. Right: Magnetic structure of the 5x5 N\'{e}el skyrmion used in the transport caluclation. }
\end{figure}

The size of skyrmions in an iron based system strongly depends on the adjacent layers. Skyrmions can be as large as 2 $ \mu$m \cite{BlwSkyr} but also shrink to a diameter of $\sim$8 atoms (2 nm). This can be achieved, for example, by putting a Pd/Fe bilayer on an Ir(111) substrate \cite{ncmr}. Thus, in general the skyrmion size can be tuned by the choice of the NM1 layer material. In our study we will show that the NM1 layer is crucial for the detection efficiency as well.
Since in this proof-of-principle study we focus on electrical detection, we choose established materials for the NM1 layer. When we perform atomistic spin dynamics simulations using parameters obtained from {\it ab initio} calculations (DMI and Heisenberg exchange are calculated using the methods \cite{ebert1} and \cite{ebert2}, respectively) for the Cu/Fe/MgO/Cu system, we get skyrmions with a diameter of approximately 50 atoms, as can be seen on the left side of figure \ref{fig:nskyr}. 
Such sizes can at the moment not be handled by {\it ab initio} methods. Therefore, we consider the artificial skyrmion structure shown on the right side of figure \ref{fig:nskyr}, which fits into a 5$\times$5 supercell and was obtained from atomistic spin dynamics simulations using artificial parameters. We inserted this skyrmion structure into all three systems.

All DFT calculations were performed with our Korringa-Kohn-Rostoker (KKR) code using a local-density approximation (LDA) exchange correlation functional \cite{lda}. The KKR code has a non-collinear fully relativistic solver with no need for perturbation theory. While self-consistent calculations utilized supercells,  transport calculations were performed using semi-infinite leads, representing a realistic transport geometry. The transport properties of the system, i.e. the transmission functions, were calculated using the Keldysh formalism, with the non-equilibrium Green's function (NEGF) as its main object \cite{negf1,negf2}.

After a number of convergence tests we decided to use an angular momentum cut-off of $l_\textrm{max}=3$ for the self-consistent calculation of the collinear magnetic structure. With these self-consistent potentials, we constructed the $5\times 5$ in-plane supercell for the non-collinear magnetic structure. Subsequent self-consistent cycles on the non-collinear structures lead only to a minor change of their properties and were thus omitted in the calculations presented in this work. Furthermore, $l_\textrm{max}=2$ was sufficient for all transport calculations. For the collinear cell we used a $540\times 540$ in-plane k-point mesh, while due to the smoothening of the transmission function by the non-collinear structure, a $18\times 18$ $k$-mesh in the supercell could be used.

For analysis of the transport states we unfolded the $k_{\parallel}$ decomposition from the Brillouin zone (BZ) of the supercell to the BZ of the primitive unit cell. This takes place similarly as described in \cite{specdiff} and provides an expression for the transmission components $T(k_\parallel,k'_ \parallel)$ non-diagonal in $k_{\parallel}$. The specular contribution was calculated as $T_{\textrm{spec}}(k_{\parallel})=T(k_\parallel,k_ \parallel)$. The diffusive contribution has two maps, $T_{\textrm{diff,in}}(k_{\parallel})$ and $T_{\textrm{diff,out}}(k_{\parallel})$, describing the distribution of tunneling electrons resolved by $k_\parallel$  within the bottom and top electrode, respectively.

\begin{align*}
T_{\textrm{diff,in}}(k_{\parallel})&=\sum_{k'_\parallel\neq k_\parallel}T(k_\parallel,k'_ \parallel)\\
T_{\textrm{diff,out}}(k_{\parallel})&=\sum_{k'_\parallel\neq k_\parallel}T(k'_\parallel,k_ \parallel)
\end{align*}

\begin{figure*}
\centering
\includegraphics[scale=0.25]{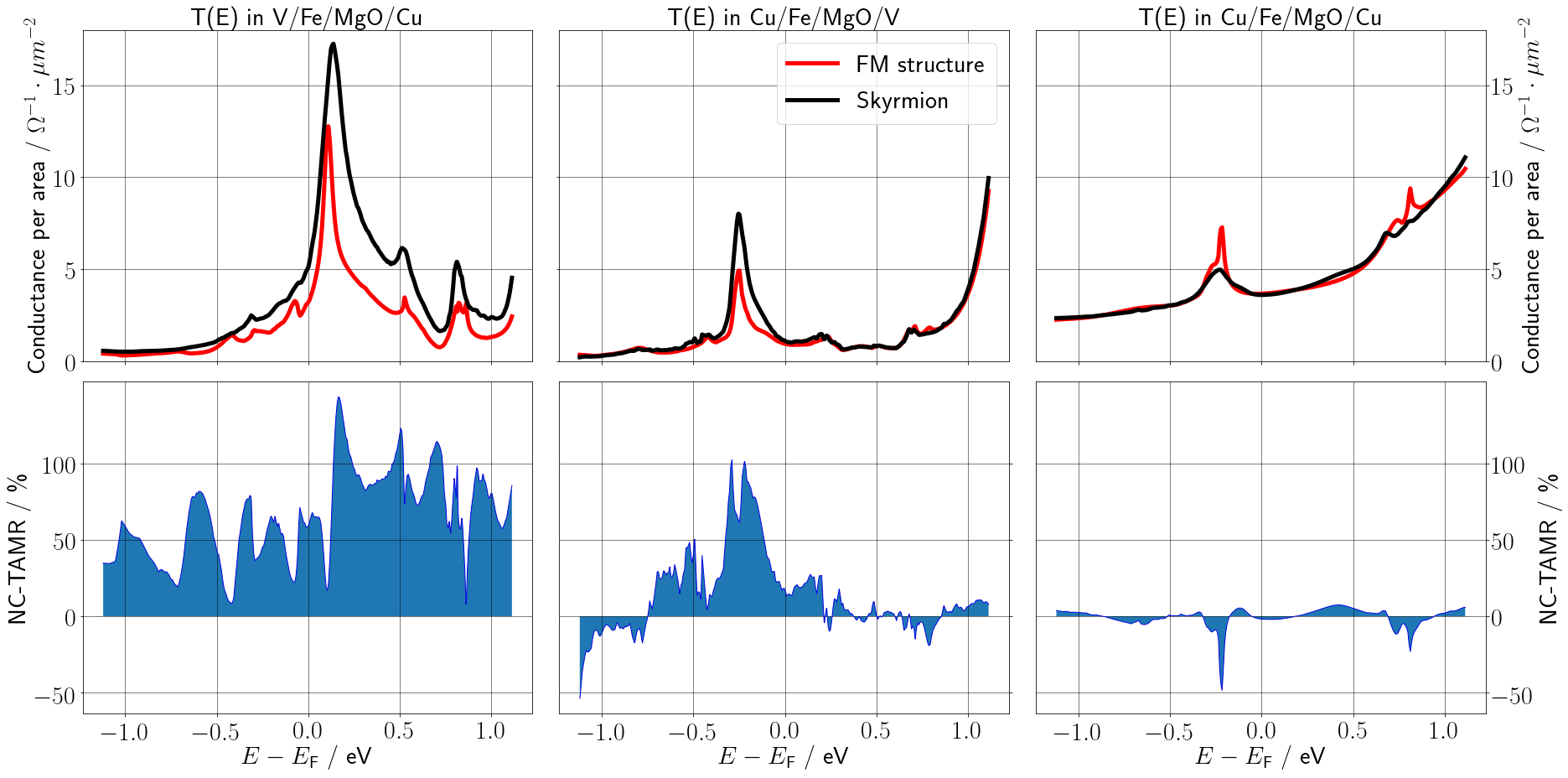}
\caption{\label{fig:TvonE} Energy dependent conductance and the resulting NCMR effect for all systems under consideration. In case of V/Fe/MgO/Cu we predict a huge effect over a wide energy range. }
\end{figure*}

The conductance of the junction is linked to the transmission $T$ via the Landauer formula $g=g_0\,T$, where $g_0$ is the conductance quantum. In the following, we analyze the energy dependent conductance. This is useful for several reasons: First, the position of the Fermi level can be tuned in experiments by alloying. Second, by applying bias voltages, states across different energies may contribute to transport. Lastly, obtaining energy-dependent transmission curves may help understand the robustness of NCMR.

In the first row of figure \ref{fig:TvonE}, the calculated energy dependent conductance curves are plotted for three material combinations, each with either a skyrmion or a purely ferromagnetic structure with out-of-plane magnetization (FM) present. The resulting NCMR ratio is the difference between the transmission curves divided by the smaller of the two values. This ratio is plotted in the second row. A clear material dependence of the NCMR ratio is apparent, where we observe the largest effect for the V/Fe/MgO/Cu system.

This behaviour must be seen in contrast to the result of \cite{lounis}, 
where the resulting TXMR effect varies sizably over the considered energy range, reaching a maximum value of about $20 \%$ at a peak 0.4\,eV away from the Fermi energy. Although we assume a skyrmion of similar size, the NCMR ratio exceeds $20 \%$ for almost all energies, reaching over 125$\%$ at its peak. Since chemical disorder at the interfaces might be able to diminish the effect, it has to be investigated in the future.

\begin{figure}
\includegraphics[scale=0.2]{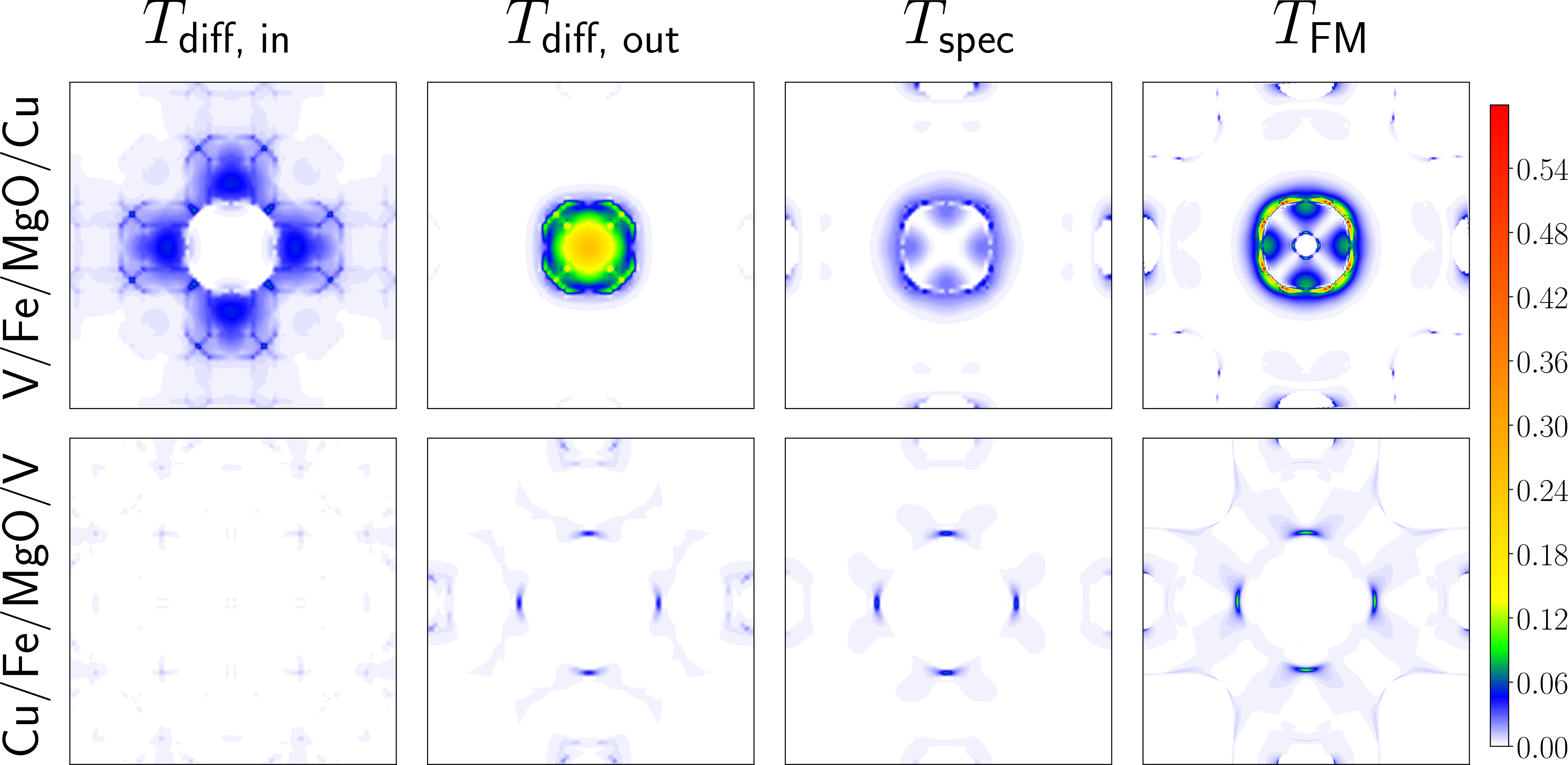}
\caption{\label{fig:TvonK_combined} $k_\parallel$-resolved transmission through V/Fe/MgO/Cu and Cu/Fe/MgO/V junctions across the BZ at the Fermi energy. The columns depict: the diffusive distributions in the bottom lead ($T_{\textrm{diff,in}}$) and in the top lead ($T_{\textrm{diff,out}}$), the specular contribution ($T_\textrm{spec}$), and transmission without skyrmion ($T_\textrm{FM})$. 
}
\end{figure}

\begin{figure*}
\includegraphics[scale=1.1]{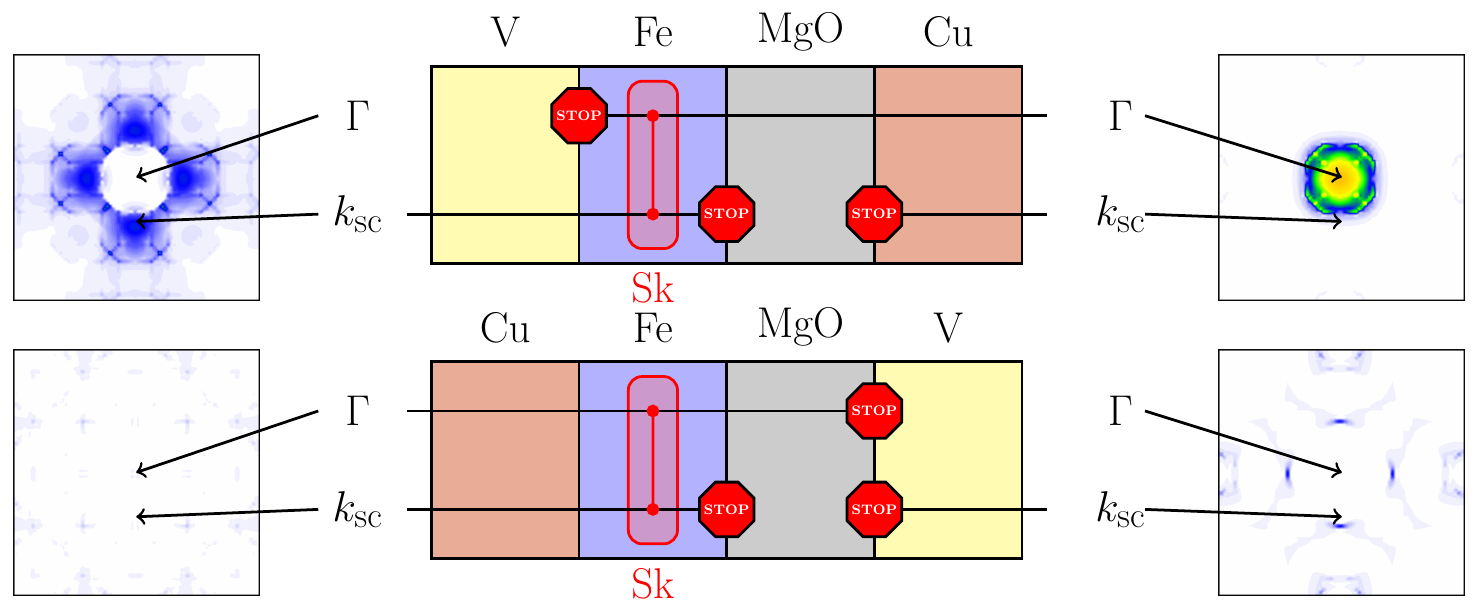}

\caption{\label{fig:mechanism} At the Fermi energy $k_\parallel$-scattering in the skyrmion (Sk) layer allows electrons coming from V to become $\Delta_1$-like and travel well through the MgO barrier. When Cu and V are switched, $k_\parallel$-scattering in the skyrmion layer becomes ineffectual.
}
\end{figure*}

The large difference between the considered systems reveals that the choice of lead materials can make a huge difference in the resulting effect. In order to get deeper insight into the mechanism behind this observation, we decomposed the transmission into its specular ($k_\parallel$ conserving) and diffusive ($k_\parallel$ non-conserving) components. This reveals the $k_\parallel$-resolved transmission unfolded into the BZ of the primitive unit cell. To keep the discussion simple, we will interpret the electrons as scattering ``in'' from the bottom electrode, transmitting first through the iron layer, next through the tunnel barrier, and then scattering ``out'' through the top lead. Of course the discussion can be equally carried out in the opposite direction.

Figure \ref{fig:TvonK_combined} displays transmittance maps for both vanadium based systems, showing the transmission for the ferromagnetic ground state with out-of-plane magnetization (FM) in the rightmost column. The other three columns show the specular contribution and the diffusive contributions referring to in- and out-scattered electrons when a skyrmion is present in the MTJ. Both transmission decompositions are performed at the respective Fermi energy. In both systems the specular contribution is mostly a scaled down version of the transmission in the ferromagnetic state. The large increase of the total transmission in case of V/Fe/MgO/Cu must be attributed to diffusive scattering. As the diffusive transmission maps show, electrons may change their $k_\parallel$-vector as they traverse the translational invariance breaking skyrmion. The mechanism is sketched in figure \ref{fig:mechanism}, where electrons with $k_\parallel$ distant ($k_{sc}$) from the $\Gamma$-point in vanadium can scatter at the skyrmion layer. After being scattered to the $\Gamma$-point, becoming $\Delta_1$-like, they can tunnel well through the MgO barrier into the copper lead. 
Note that the vanadium does not provide any incoming states at the $\Gamma$-point suitable for crossing the MgO barrier.
This means that the system essentially acts as a spin valve with only one magnetic layer which can be opened using a skyrmion as a switch.

The same simple model also explains why the total effect in the Cu/Fe/MgO/V system is relatively suppressed at the Fermi energy. 
While the skyrmion may still scatter states between the $\Gamma$-point and $k_{\text{sc}}$, any $\Delta_1$-like electrons crossing the barrier cannot be injected into the vanadium lead.

The relationship between the mechanism reported in references \cite{lounis,ncmr,hann} and the mechanism in our study is thus comparable to the difference between TMR in SP-STM and TMR in MgO based barriers, where the latter is also much larger due to high $k_{\parallel}$ polarization with symmetry filtering in MgO \cite{butler2008tunneling,heiliger2006microscopic}. Since the mechanism in our study goes beyond simple spin-mixing and utilizes the MgO symmetry filtering, we call the effect NCMR \cite{ncmr,hann}, which is the more general name. We could thereby explain that the nature of the non-collinear contribution can vary strongly from system to system, and that the choice of lead materials impacts the effect size. For V/Fe/MgO/Cu we predict a large NCMR ratio exceeding $125\%$ at its peak. 
Even though this is a proof-of-principle study working with small skyrmions, the large NCMR ratios calculated for V/Fe/MgO/Cu motivate further research. With heavy metal underlayers in future works for which such small skyrmions are stable, similarly high effects might be achievable.

\FloatBarrier

\bibliographystyle{h-physrev}

\bibliography{research}

\end{document}